# Analytical Formulae for the Loss Factors and Wakefields of a Rectangular Accelerating Structure

J. Gao, Z.C. Liu

Institute of High Energy Physics (IHEP), Beijing 100049, China

**Abstract:** Wakefields in a rectangular accelerating structure can be calculated in time domain by directly solving Maxwell's equations by a 3D code. In this paper, we will give analytical formulae to calculate the synchronous modes' loss factors. From these analytical formulae on can get the delta function wakefields. The relations between the loss factors (wakefields) and the structure geometrical dimensions are well established. These analytical expressions of loss factors can be used also in a single rectangular resonant cavity. It is shown that the potential application of a rectangular accelerating structure is to accelerate a flat beam in a linear collider.

**1 Introduction**

In this paper, we will consider the rectangular slow wave accelerating structure shown in Fig. 1. Since this is a 3D problem, usually, one has to resort to a 3D code to calculate the wakefields excited inside the structure by a passing charge. In ref. 1, the analytical formulae of the loss factors and the wakefields in a cylindrical disk-loaded accelerating structure have been established. By using the same method as in ref. 1, we will give the analytical formulae of the loss factors and wakefields in a rectangular slow wave accelerating structure.

In section 2, the single rectangular cavity's properties are briefly summarized, since a rectangular cavity is the very basic element of a rectangular accelerating structure. In section 3, the analytical expressions for all the synchronous modes' loss factors and the wakefields are derived. In section 4, two examples are calculated analytically, one is a closed rectangular cavity and another is a rectangular slow wave accelerating structure. Finally, some conclusions are drawn in section 5.

**2 Single Rectangular Resonant Cavity**

In the cartesian coordinate system the *em* field distributions of the TM$_{mnl}$ modes in a rectangular resonant cavity are:

$$E_x = -\frac{H_0 l\pi m\pi}{j\omega_{mnl}\varepsilon_0 hb}\cos\left(\frac{m\pi x}{b}\right)\sin\left(\frac{n\pi y}{a}\right)\sin\left(\frac{l\pi z}{h}\right) \qquad (1)$$

$$E_y = -\frac{H_0 l\pi n\pi}{j\omega_{mnl}\varepsilon_0 ha}\sin\left(\frac{m\pi x}{b}\right)\cos\left(\frac{n\pi y}{a}\right)\sin\left(\frac{l\pi z}{h}\right) \qquad (2)$$

$$E_z = \frac{H_0}{j\omega_{mnl}\varepsilon_0}\left(\left(\frac{m\pi}{b}\right)^2+\left(\frac{n\pi}{a}\right)^2\right)\sin\left(\frac{m\pi x}{b}\right)\sin\left(\frac{n\pi y}{a}\right)\sin\left(\frac{l\pi z}{h}\right) \qquad (3)$$

$$H_x = \frac{H_0 n\pi}{a} \sin\left(\frac{m\pi x}{b}\right) \cos\left(\frac{n\pi y}{a}\right) \cos\left(\frac{l\pi z}{h}\right) \qquad (4)$$

$$H_y = -\frac{H_0 m\pi}{b} \cos\left(\frac{m\pi x}{b}\right) \sin\left(\frac{n\pi y}{a}\right) \cos\left(\frac{l\pi z}{h}\right) \qquad (5)$$

$$H_z = 0 \qquad (6)$$

$$m = 1, 2, \cdots, n = 1, 2, \cdots, l = 0, 1, 2, \cdots \qquad (7)$$

Where a, b and h are shown in Fig. 1, $\varepsilon_0$ is the electric permittivity in vacuum, and $H_0$ is a constant.

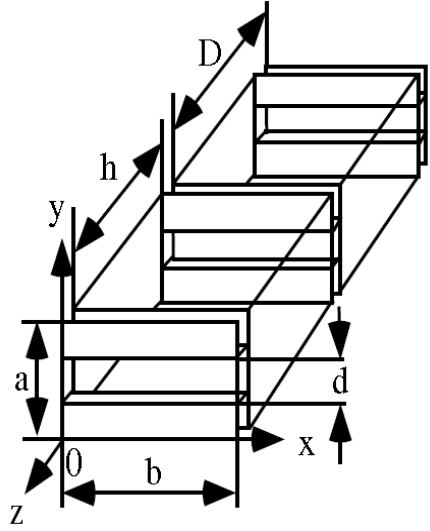

Figure 1: A rectangular accelerating structure

The resonant angular frequencies of the TM$_{mnl}$ modes are determined by:

$$\omega_{mnl} = c\left((m\pi/b)^2 + (n\pi/a)^2 + (l\pi/h)^2\right)^{1/2} \qquad (8)$$

The power dissipation $P_{mnl}$, stored energy $U_{mnl}$, and quality factor $Q_{0,mnl}$ are expressed as [2]:

$$P_{mnl} = \frac{R_{s,mnl} H_0^2 \pi^2}{4} \left(\frac{n^2 b}{a}\left(1 + \frac{h}{a\delta}\right) + \frac{m^2 a}{b}\left(1 + \frac{h}{b\delta}\right)\right) \qquad (9)$$

$$U_{mnl} = \frac{H_0^2 \mu_0 abh\pi^2}{16\delta}\left(\frac{n^2}{a^2} + \frac{m^2}{b^2}\right) \qquad (10)$$

$$Q_{0,mnl} = \frac{\omega \mu_0 abh\left(n^2/a^2 + m^2/b^2\right)}{4R_{s,mnl}\delta\left(n^2 \frac{b}{a}(1 + h/a\delta) + m^2 \frac{a}{b}(1 + h/b\delta)\right)} \qquad (11)$$

Where

$$\delta = \begin{cases} 1, l \neq 0 \\ 1/2, l = 0 \end{cases} \qquad (12)$$

$$R_{s,mnl} = \left(\frac{\omega_{mnl}\mu_0}{2\sigma}\right)^{1/2} \tag{13}$$

Where $\sigma$ is the electric conductivity and $\mu_0$ is the magnetic permeability, respectively.

**3 Loss Factors of a Rectangular Slow Wave Accelerating Structure**

We consider a rectangular accelerating structure as shown in Fig. 1. Due to the coupling through the aperture between cavities, passbands will form corresponding to each rectangular resonant mode. We will keep using three subscripts *mnl* to specify the passbands. In the following we assume that the synchronous frequency of the TM$_{mnl}$ passband is not very different from $\omega_{mnl}$, and we will use $\omega_{mnl}$ to replace the corresponding synchronous frequency. The delta function wakefields of a point charge passing through a rectangular structure can be found by using the following formulae

$$W_x(\tau) = \sum_{m=0}^{\infty}\sum_{n=1}^{\infty}\sum_{l=0}^{\infty} W_{x,mnl}(\tau) \tag{14}$$

$$W_y(\tau) = \sum_{m=0}^{\infty}\sum_{n=1}^{\infty}\sum_{l=0}^{\infty} W_{y,mnl}(\tau) \tag{15}$$

$$W_z(\tau) = \sum_{m=0}^{\infty}\sum_{n=1}^{\infty}\sum_{l=0}^{\infty} W_{z,mnl}(\tau) \tag{16}$$

Where $W_{x,mnl}$, $W_{y,mnl}$ and $W_{z,mnl}$ are the wakefields corresponding to the *mnl*th synchronous mode. To find out the expressions of $W_{x,mnl}$, $W_{y,mnl}$ and $W_{z,mnl}$, one has to use the generalized Panofsky-Wenzel theorem derived in ref. 3. We know therefore that in a cartesian coordinate system

$$W_{x,mnl}(s) = Z_l(s)\frac{\partial T_{mn}(x,y)}{\partial x} \tag{17}$$

$$W_{y,mnl}(s) = Z_l(s)\frac{\partial T_{mn}(x,y)}{\partial y} \tag{18}$$

$$W_{z,mnl}(s) = T_{mn}(x,y)\frac{dZ_l(s)}{ds} \tag{19}$$

Where $s=\tau c$, $c$ is the velocity of light in vacuum and $s$ is the distance between the exciting charge and a test charge. $T_{mn}(x,y)$ and $Z_l(s)$ satisfy the following equations:

$$Z_l(s)\frac{\partial^2 T_{mn}(x,y)}{\partial x^2} + Z_l(s)\frac{\partial^2 T_{mn}(x,y)}{\partial y^2} - T_{mn}(x,y)\frac{d^2 Z_l(s)}{dz^2} = 0 \tag{20}$$

It is found that

$$W_{z,mnl}(\tau) = 2k_{mnl}\frac{\sin\left(\frac{m\pi x}{b}\right)\sin\left(\frac{n\pi y}{a}\right)}{\sin\left(\frac{m\pi x_w}{b}\right)\sin\left(\frac{n\pi y_w}{a}\right)} \times \frac{\sin\left(\frac{m\pi x_q}{b}\right)\sin\left(\frac{n\pi y_q}{a}\right)}{\sin\left(\frac{m\pi x_w}{b}\right)\sin\left(\frac{n\pi y_w}{a}\right)}\cos(\omega_{mnl}\tau)$$

$$W_{x,mnl}(\tau) = 2m \frac{c\pi k_{mnl}}{\omega_{mnl} b} \frac{\cos\left(\frac{m\pi x}{b}\right)\sin\left(\frac{n\pi y}{a}\right)}{\sin\left(\frac{m\pi x_w}{b}\right)\sin\left(\frac{n\pi y_w}{a}\right)} \times \frac{\sin\left(\frac{m\pi x_q}{b}\right)\sin\left(\frac{n\pi y_q}{a}\right)}{\sin\left(\frac{m\pi x_w}{b}\right)\sin\left(\frac{n\pi y_w}{a}\right)} \sin(\omega_{mnl}\tau) \quad (22)$$

$$W_{y,mnl}(\tau) = 2n \frac{c\pi k_{mnl}}{\omega_{mnl} a} \frac{\sin\left(\frac{m\pi x}{b}\right)\cos\left(\frac{n\pi y}{a}\right)}{\sin\left(\frac{m\pi x_w}{b}\right)\sin\left(\frac{n\pi y_w}{a}\right)} \times \frac{\sin\left(\frac{m\pi x_q}{b}\right)\sin\left(\frac{n\pi y_q}{a}\right)}{\sin\left(\frac{m\pi x_w}{b}\right)\sin\left(\frac{n\pi y_w}{a}\right)} \sin(\omega_{mnl}\tau) \quad (23)$$

Where $x_q$ and $y_q$ are the transverse coordinates of the exciting charge, $x$ and $y$ are the transverse coordinates of the test charge, and the axis of $x=x_w$, $y=y_w$ has been chosen to be on the surface of the waveguide which connects the two adjacent rectangular cavities.

The definition of the loss factor of a synchronous mode is expressed as [4]

$$k_{mnl} = \frac{E_{s,z}^{mnl}(x=x_w, y=y_w)^2}{4dU_{mnl}/dz} \quad (24)$$

Where $E_{s,z}^{mnl}(x=x_w, y=y_w)$ is the synchronous decelerating electric field along the axis of $x=x_w$, $y=y_w$, and $dU_{mnl}/dz$ is the energy stored per meter. For the *mnl*th passband by using the same method as in refs. 1 and 5, one has $E_{s,z}^{mnl}(x=x_w, y=y_w) = E_{z,mnl}(x=x_w, y=y_w)\eta(\theta_{mnl})$ and $dU_{mnl}/dz = U_{mnl}/D$, where $E_{z,mnl}(x=x_w, y=y_w)$ is the longitudinal electric field of the *mnl*th mode in the rectangular cavity before the apertures are opened. When $l=0$,

$$\eta(\theta_{mn0})^+ = \frac{\sin(\theta_{mn0}h/2D)}{\theta_{mnl}} \quad (25)$$

and when $l \neq 0$, one has two synchronous modes corresponds to the indice $l$ as explained in ref. 1

$$\eta(\theta_{mnl})^+ = \frac{1}{2}\frac{\sin(\theta_{mnl}h/2D + l\pi/2)}{\theta_{mnl} + lD\pi/h} \quad (26)$$

and

$$\eta(\theta_{mnl})^- = \frac{1}{2}\frac{\sin(\theta_{mnl}h/2D - l\pi/2)}{\theta_{mnl} - lD\pi/h} \quad (27)$$

where

$$\theta_{mnl} = D\left(\left(\frac{m\pi}{b}\right)^2 + \left(\frac{n\pi}{a}\right)^2 + \left(\frac{l\pi}{h}\right)^2\right)^{1/2} \quad (28)$$

By using eq. 3, eq. 10, eq. 24 and eq. 26, we get the general expression of the loss factor $k_{mnl}$ corresponding to the $mnl$th passband

$$k_{mnl} = \frac{E_{s,z}^{mnl}(x=x_w, y=y_w)^2 D}{4U_{mnl}}$$

$$= \frac{4h\left((m\pi/b)^2 + (n\pi/a)^2\right)\sin^2\left(\frac{m\pi x_w}{b}\right)\sin^2\left(\frac{n\pi y_w}{a}\right)}{\varepsilon_0 abD\left((m\pi/b)^2 + (n\pi/a)^2 + (l\pi/h)^2\right)}\left(\frac{S(x_1)^2 + S(x_2)^2}{2}\right) \quad (29)$$

$$= k_{mnl}^* \sin^2\left(\frac{m\pi x_w}{b}\right)\sin^2\left(\frac{n\pi y_w}{a}\right)$$

where

$$k_{mnl}^* = \frac{4h\left((m\pi/b)^2 + (n\pi/a)^2\right)}{\varepsilon_0 abD\left((m\pi/b)^2 + (n\pi/a)^2 + (l\pi/h)^2\right)}\left(\frac{S(x_1)^2 + S(x_2)^2}{4}\right) \quad (30)$$

$$S(x) = \frac{\sin(x)}{x} \quad (31)$$

and

$$x_1 = \frac{h}{2}\left(\left(\left(\frac{m\pi}{b}\right)^2 + \left(\frac{n\pi}{a}\right)^2 + \left(\frac{l\pi}{h}\right)^2\right)^{1/2} - \frac{l\pi}{h}\right) \quad (32)$$

$$x_2 = \frac{h}{2}\left(\left(\left(\frac{m\pi}{b}\right)^2 + \left(\frac{n\pi}{a}\right)^2 + \left(\frac{l\pi}{h}\right)^2\right)^{1/2} + \frac{l\pi}{h}\right) \quad (33)$$

By inserting eq. 29 into eqs. 21, 22 and 23, one finds finally

$$W_{z,mnl}(\tau) = 2k_{mnl,i}\sin\left(\frac{m\pi x}{b}\right)\sin\left(\frac{n\pi y}{a}\right) \times \sin\left(\frac{m\pi x_q}{b}\right)\sin\left(\frac{n\pi y_q}{a}\right)\cos(\omega_{mnl}\tau) \quad (34)$$

$$W_{x,mnl}(\tau) = 2m\frac{c\pi k_{mnl,i}}{\omega_{mnl}b}\cos\left(\frac{m\pi x}{b}\right)\sin\left(\frac{n\pi y}{a}\right) \times \sin\left(\frac{m\pi x_q}{b}\right)\sin\left(\frac{n\pi y_q}{a}\right)\sin(\omega_{mnl}\tau)$$

(35)

$$W_{y,mnl}(\tau) = 2n\frac{c\pi k_{mnl,i}}{\omega_{mnl}a}\sin\left(\frac{m\pi x}{b}\right)\cos\left(\frac{n\pi y}{a}\right) \times \sin\left(\frac{m\pi x_q}{b}\right)\sin\left(\frac{n\pi y_q}{a}\right)\sin(\omega_{mnl}\tau)$$

(36)

where the subscript $i$ distinguishes four different cases shown in Fig. 2, and $k_{mnl,i}$ are expressed as

$$k_{mnl,1} = k_{mnl}^* \sin^2\left(\frac{n\pi y_w}{a}\right) \quad (37)$$

$$k_{mnl,2} = k^*_{mnl} \sin^2\left(\frac{m\pi x_w}{b}\right) \tag{38}$$

$$k_{mnl,3} = k^*_{mnl} \sin^2\left(\frac{n\pi y_w}{a}\right)\sin^2\left(\frac{m\pi x_w}{b}\right) \tag{39}$$

$$k_{mnl,4} = k^*_{mnl} \tag{40}$$

The manually added aperture dependent coefficients in Eqs.37 to 39 simply represent the influence of the coupling aperture on the loss factors. It is obvious that all the loss factors will be zero when $x_w=0$ and $y_w=0$, since this case corresponds to an infinite uniform rectangular waveguide (assuming this waveguide has no loss). It is important to note that the dependence of the wakefields on the transverse charge coordinates in a rectangular structure is totally different from that in a cylindrical structure, and it is this difference which implies the potential application of rectangular accelerating structure in future linear colliders.

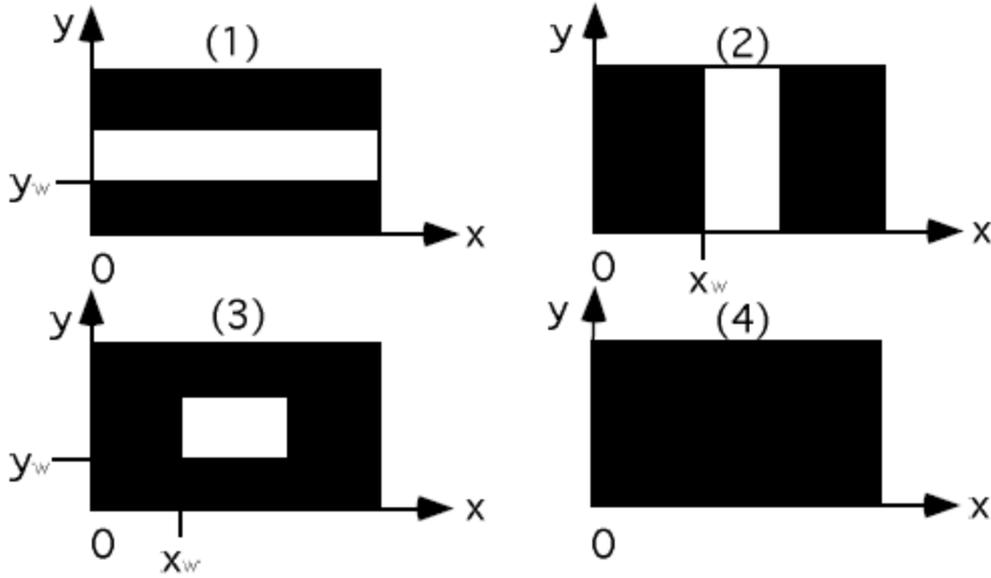

Figure 2: Four types of coupling apertures.

By setting $D=h$ and using eq. 39 one gets the wakefields in a closed rectangular resonant cavity. For a Gaussian bunch of charge $q$ and bunch length $\sigma_t$ one can calculate the integrated wakefield started from the delta function wakefields:

$$W_{G,z}(\tau) = \int_{-\infty}^{\tau} W_z(\tau-t)I(t)dt \tag{41}$$

$$W_{G,x}(\tau) = \int_{-\infty}^{\tau} W_x(\tau-t)I(t)dt \tag{42}$$

$$W_{G,y}(\tau) = \int_{-\infty}^{\tau} W_y(\tau-t)I(t)dt \tag{43}$$

where

$$I(t) = \frac{q}{(2\pi)^{1/2} \sigma_t} \exp\left(-\frac{t^2}{2\sigma_t^2}\right) \tag{44}$$

If $\tau \geq 3\sigma_t$ eqs. 41, 42 and 43 can be replaced by the following expressions:

$$W_{G,z}(\tau) = \sum_{m=0}^{\infty}\sum_{n=1}^{\infty}\sum_{l=0}^{\infty} W_{z,mnl}(\tau) \exp\left(-\frac{\omega_{mnl}^2 \sigma_t^2}{2}\right) \tag{45}$$

$$W_{G,x}(\tau) = \sum_{m=0}^{\infty}\sum_{n=1}^{\infty}\sum_{l=0}^{\infty} W_{x,mnl}(\tau) \exp\left(-\frac{\omega_{mnl}^2 \sigma_t^2}{2}\right) \tag{46}$$

$$W_{G,y}(\tau) = \sum_{m=0}^{\infty}\sum_{n=1}^{\infty}\sum_{l=0}^{\infty} W_{y,mnl}(\tau) \exp\left(-\frac{\omega_{mnl}^2 \sigma_t^2}{2}\right) \tag{47}$$

At the end of this section we will discuss the focusing properties of the fundamental mode. Different from a circular structure where the fundamental accelerating mode electromagnetic field does not change the transverse momentum of a charged particle travelling at the speed of light (almost), the fundamental accelerating mode in a rectangular structure focuses or defocuses the passing charged particle towards or from the axis of the structure depending on the phase of the accelerated particle with respect to the accelerating electric field. This characteristic property makes a rectangular structure much more interesting for future linear colliders where the beam should be kept very close to the axis to avoid the excessive wakefield deflecting forces. The expressions of the forces in the three directions can be obtained easily from eqs. 34 to 36

$$F_z(\Phi) = qE_0 \cos(\Phi) \tag{48}$$

$$F_x(\Phi) = -qE_0 \frac{c\pi}{\omega_{110} b} \cos\left(\frac{\pi x}{b}\right) \sin\left(\frac{\pi y}{a}\right) \sin(\Phi) \tag{49}$$

$$F_y(\Phi) = -qE_0 \frac{c\pi}{\omega_{110} a} \sin\left(\frac{\pi x}{b}\right) \cos\left(\frac{\pi y}{a}\right) \sin(\Phi) \tag{50}$$

where $q$ is the charge of the particle, $E_0$ is the synchronous accelerating electric field strength, and $\Phi > 0$ corresponds to the particle located before the synchronous rf crest. If the particle is located close to the axis, one has

$$F_x(\Phi) = qE_0 \frac{c\pi^2}{\omega_{110} b^2} dX \sin(\Phi) \tag{51}$$

$$F_y(\Phi) = qE_0 \frac{c\pi^2}{\omega_{110} a^2} dY \sin(\Phi) \tag{52}$$

where $dX$ and $dY$ are the particle's transverse coordinates measured from the center of the rectangular structure. It is clear that by choosing negative $\Phi$ (where BNS damping works [6]) one can get focusing in both transverse planes.

## 4 Example and Discussion

In this section, we will give one example to show the analytical results. The loss factor which will be plotted in the following figures is defined as $k_{mnl,i}(\sigma_z) = k_{mnl,i} \exp(-\omega_{mnl}^2 \sigma_z^2/c^2)$, where $\sigma_z = c\sigma_t$, $m$=1 to 15, $n$=1 to 15 and $l$=1 to 10.

We calculated the loss factors and the wakefields in a closed rectangular cavity driven by a bunch of a Gaussian charge distribution by formulae and compared with CST-MWS simulation results [7]. Taking $a$=6 cm, $b$=9 cm, $h$=$D$=2.92 cm, $x_q$=$x$=$b/2$, $y_q$=$y$=$a/2$, and $\sigma_z$=2.5 mm, one finds $k_{mnl,4}(\sigma_z)$ vs resonant frequency and $W_z(\sigma_z)$ shown in Fig. 3 calculated by using eqs. 45 and $W_z(\sigma_z)$ simulated by CST-MWS. $W_x(\sigma_z)$ and $W_y(\sigma_z)$ are shown in Fig.4 calculated by eqs. 46 and 47 with $x_q$=$x$=$b/2$+0.8 cm, $y_q$=$y$=$a/2$+0.8 cm, and $\sigma_z$=2.5 mm and simulated by CST-MWS separately.

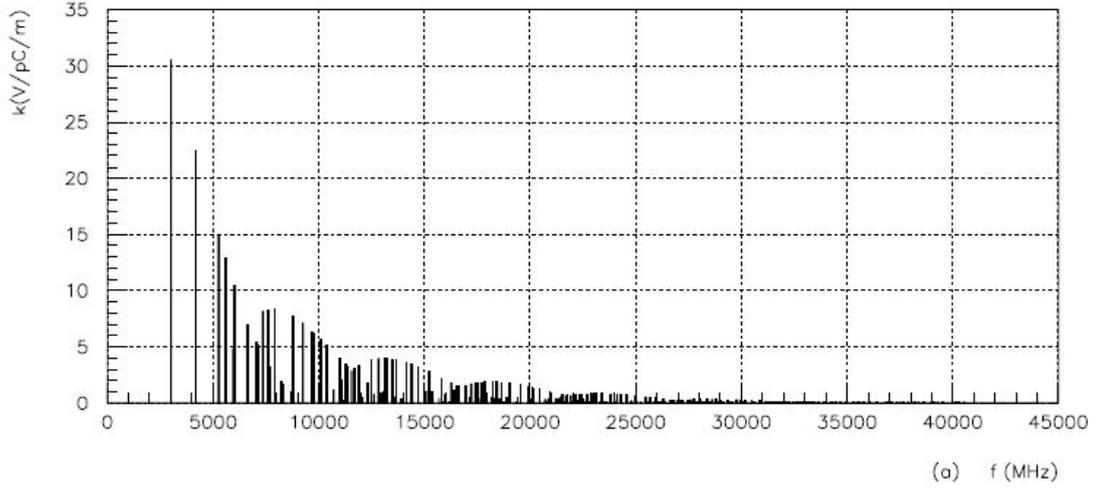

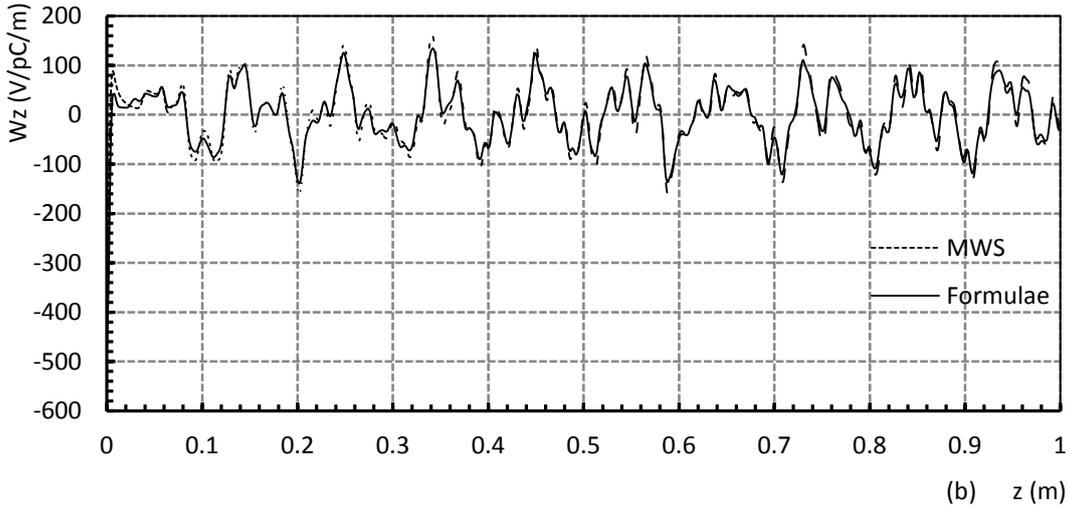

Figure 3: (a) Loss factors $k_{mnl,4}(\sigma_z)$ vs frequency, and (b) $W_{G,z}$ (V/pC/m) vs distance calculated by formulae and CST-MWS. For both figures $\sigma_z$=2.5 mm, $x$=$x_q$= $b/2$, and $y$=$y_q$= $a/2$. The dimension of the structure: $D$=$h$=2.92 cm, $a$=6 cm, and $b$=9 cm.

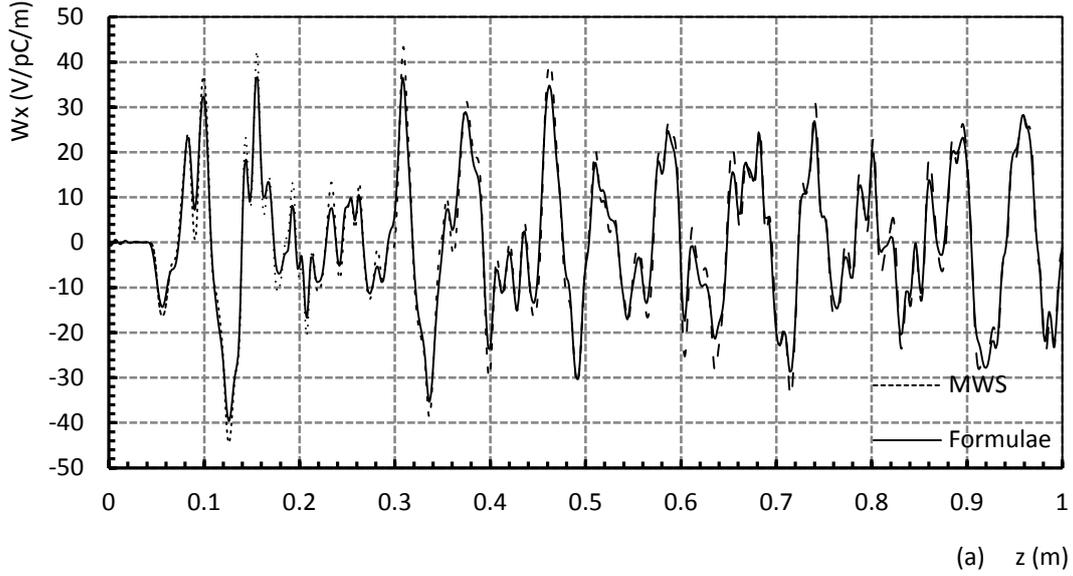

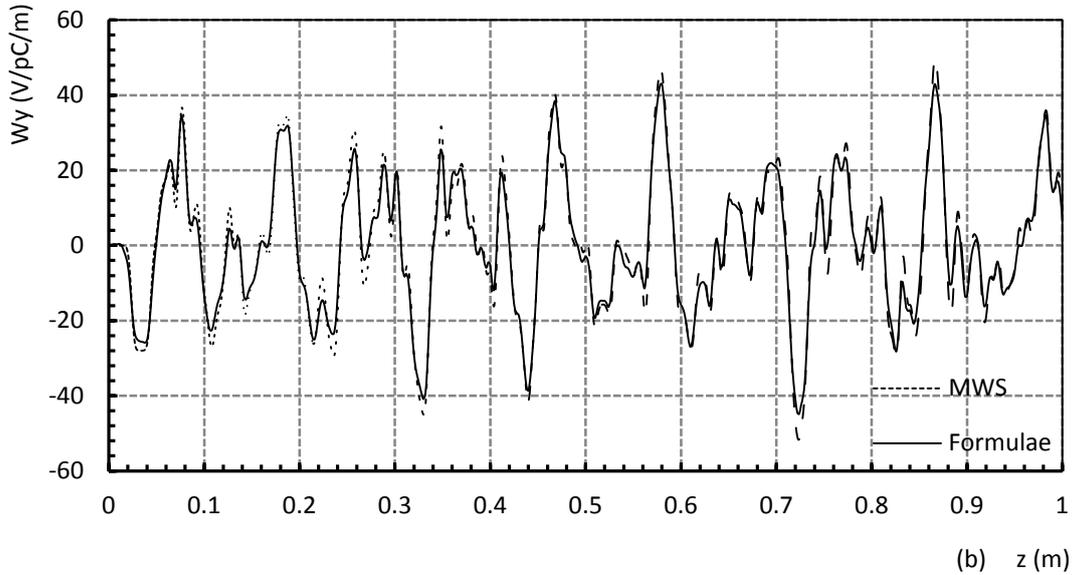

Figure 4: (a) $W_{G,x}$ (V/pC/m) vs distance calculated by formulae and CST-MWS, and (b) $W_{G,y}$ (V/pC/m) vs distance calculated by formulae and CST-MWS. For both figures $\sigma_z$=2.5 mm, $x=x_q= b/2+0.8$ cm, and $y=y_q= a/2+0.8$ cm. The dimension of the structure: D=h=2.92 cm, a=6 cm, and b=9 cm.

**5 Conclusions**

The analytical formulae of the loss factors and the wakefields in a closed rectangular cavity and in a slow wave rectangular accelerating structure have been derived. One illustrative example have been given to show the wakefields excited in these cavities working at S-band frequency (the frequency of the fundamental mode has been chosen around 3GHz). The formulae calculation results agree well with the CST-MWS simulation results.

The method used in this paper and in ref. 1 can be easily applied to the elliptical cavity or spherical cavity cases where the corresponding analytical electromagnetic field solutions exist.

## 6 Acknowledgements

We acknowledge the support of National Natural Science Foundation of China (NSFC, Project 11175192 and 11275226).